\documentclass[utf8]{frontiersSCNS} 
\usepackage{url,lineno,microtype,subcaption}
\usepackage[onehalfspacing]{setspace}
\usepackage{multirow}
\usepackage{booktabs}

\def\keyFont{\fontsize{8}{11}\helveticabold }
\def\firstAuthorLast{Zeng {et~al.}} 
\def\Authors{Zheni Zeng\,$^{1,\dagger}$, 
Chaojun Xiao\,$^{1,\dagger}$, 
Yuan Yao\,$^{1}$, 
Ruobing Xie\,$^{2}$, 
Zhiyuan Liu\,$^{1,\ast}$, 
Fen Lin\,$^{2}$, 
Leyu Lin\,$^{2}$ 
and Maosong Sun\,$^{1}$} 
\def\Address{$^{1}$Department of Computer Science and Technology
Institute for Artificial Intelligence, Tsinghua University, Beijing, China \\
\ Beijing National Research Center for Information Science and Technology, China \\
$^{2}$Search Product Center, WeChat Search Application Department, Tencent, China
}
\def\corrAuthor{Zhiyuan Liu}

\def\corrEmail{liuzy@tsinghua.edu.cn}

\begin{document}
\onecolumn
\firstpage{1}

\title[Pre-training in Recommendation]{Knowledge Transfer via Pre-training for Recommendation: A Review and Prospect
} 

\author[\firstAuthorLast ]{\Authors} 
\address{} 
\correspondance{} 

\extraAuth{}

\maketitle

\newcommand\blfootnote[1]{%
  \begingroup
  \renewcommand\thefootnote{}\footnote{#1}%
  \addtocounter{footnote}{-1}%
  \endgroup
}

\begin{abstract}

Recommender systems aim to provide item recommendations for users, and are usually faced with data sparsity problem (e.g., cold start) in real-world scenarios. Recently pre-trained models have shown their effectiveness in knowledge transfer between domains and tasks, which can potentially alleviate the data sparsity problem in recommender systems. In this survey, we first provide a review of recommender systems with pre-training. In addition, we show the benefits of pre-training to recommender systems through experiments. Finally, we discuss several promising directions for future research for recommender systems with pre-training.

\tiny
 \keyFont{ \section{Keywords:} Recommender System, Pre-trained Model, Knowledge Transfer, Cross-domain Transfer, Cold Start} 
\end{abstract}


\section{Introduction}

With the rapid development of the Internet, users are faced with information overload, where the large quantity of online items makes it hard for users to make decisions effectively. Recommender systems aim to provide recommendations by capturing user preference for items (e.g., movies, books, songs, news, websites) from explicit item ratings given by users, or implicit user-item interactions (e.g., browsing and purchasing histories). The application of recommender systems has enabled personalized services in many scenarios, such as e-commerce and website browsing. \blfootnote{$\dagger$ Indicates equal contribution. The order is determined by dice rolling.}
\blfootnote{Preprint. This paper is submitted to \textit{Frontiers in Big Data} and is under review.}

Recommender systems are usually faced with data sparsity in real-world scenarios. Recommender systems can suffer when providing recommendations for new items or users due to lack of information, which is known as the cold start problem~
\citep{8229786}. It has been shown that the data sparsity problem in recommender systems can be alleviated by transferring knowledge from other domains or tasks~\citep{cantador2015cross}, and integrating heterogeneous external knowledge~\citep{guo2020survey}. 

In the field of natural language processing, pre-trained models have achieved great success recently on a broad variety of tasks by knowledge transfer~\citep{xipengpre}. Models are usually first pre-trained on large-scale unsupervised data to learn universal language representations and then fine-tuned on downstream tasks to achieve knowledge transfer. The models can be pre-trained to learn either shallow context-free word embeddings~\citep{mikolov2013distributed}, or deep context-aware language representations~\citep{devlin2019bert}. The resulting language representations have proven to be useful not only for different tasks (e.g., natural language inference and question answering~\citep{devlin2019bert}), but also for different scenarios, such as few-shot learning~\citep{brown2020language} and domain adaptation~\citep{rietzler2020adapt}.

In the context of recommender systems, we can group works which utilize pre-training mechanism to improve the precision of recommendation into two categories: feature-based models and fine-tuning models. The feature-based models generally use pre-trained models to obtain features from side-information (e.g., the content of items and knowledge graphs) for users and items~\citep{guo2020survey}. And the fine-tuning models leverage the user-item interaction records to pre-train a deep transferable neural model, which is subsequently fine-tuned on downstream recommendation tasks~\citep{chen2019bert4sessrec}.
Generally, the benefits of pre-training to recommender systems can be summarized as twofold: (1) Pre-training tasks can better exploit user-item interaction data to capture user interests. (2) Pre-training can help integrate knowledge from different tasks and sources into universal user/item representations, which can be further adapted to various scenarios in recommender systems, such as cold start and cross-domain transfer.

The contributions of this survey can be summarized as follows: (1) \textit{Systematic Review.} We provide a systematic review of the pre-training methods for recommender systems with a clear taxonomy. (2) \textit{Empirical Results.} We present empirical results to show the benefits of pre-training to recommender systems. We conduct experiments on the task of movie recommendation where different types of knowledge are integrated by pre-training for better recommendations, especially in the cold start and cross-domain transfer scenarios. (3) \textit{Future Directions.} Based on the review and experiments, we discuss several promising directions for future research, including how to better leverage pre-training approaches to improve recommender systems, and how recommender systems can motivate better pre-training models.

The rest of the survey is organized as follows: In Section~\ref{sec:feature-model} and Section~\ref{sec:fine-model}, we provide a review of existing methods of recommender systems with pre-training. In Section~\ref{sec:experiments}, we conduct experiments to empirically show the benefits of pre-training to recommender systems. In Section~\ref{sec:outlook}, we discuss promising directions for future research.





\section{Feature-based Models}
\label{sec:feature-model}
Feature-based models leverage side-information (e.g., contents of items, knowledge graphs and social networks) using pre-trained models to directly enrich the representations of users or items. 
Different from collaborative filtering (CF) methods that learn the representations from user-item interaction records, feature-based models focus on extracting widely applicable features from external information sources with pre-trained models, and then integrate these features into the recommendation framework. By combining rich side-information and user-item interaction data, feature-based models can potentially solve some challenges such as the data sparsity problem.

The general idea can be illustrated as follows. Given the external information resource, the pre-trained models are applied to obtain the external feature vectors, $\hat{\mathbf{u}}_i$ and $\hat{\mathbf{v}}_j$ for user $u_i$ and item $v_j$ respectively. Denote $\tilde{\mathbf{u}}_i$ and $\tilde{\mathbf{v}}_j$ as the features learned from the user-item interaction records for user $u_i$ and item $v_j$ respectively. Then the final representations $\mathbf{u}_i$, $\mathbf{v}_j$ for each user and item are obtained by aggregating external feature vectors and the vectors from the user-item interaction data:
\begin{equation}
    \mathbf{u}_i = g_u(\tilde{\mathbf{u}}_i, \hat{\mathbf{u}}_i), \quad
    \mathbf{v}_i = g_v(\tilde{\mathbf{v}}_i, \hat{\mathbf{v}}_i),
    \label{eq:aggregate}
\end{equation}
where $g_u(\cdot)$ and $g_i(\cdot)$ are aggregate functions. The preference score for user $u_i$ and item $v_j$ is calculated by:
\begin{equation}
    s(u_i, v_j) = f(\mathbf{u}_i, \mathbf{v}_j),
    \label{eq:score}
\end{equation}
where $f(\cdot)$ is a recommendation function, which can be factorization machines~\citep{rendle2010factorization} and deep neural networks~\citep{he2016deep,fan2018deep}, etc. 

According to the type of external information resources, feature-based pre-trained models can be roughly categorized into content-based recommendation, knowledge graph-based recommendation, and social recommendation models. Different types of meta-information require different pre-trained models. We will introduce how the pre-training mechanism is used in these three kinds of recommender systems in detail.

\subsection{Content-based Recommendation}
Content-based recommendation assumes that users prefer items that are similar to their historical interacted items. Therefore, it is important for content-based recommender systems to encode the content of items into expressive low-dimensional representations. The pre-trained models have proven to be powerful in extracting generally applicable representations from text, images, and audio etc. Hence, many works learn features from the content of items to serve recommendation models. 

\citet{liang2015content} pre-train a multi-layer neural network to extract audio features for music recommendation via a semantic tag prediction task. In terms of dealing with textual data for recommendation, such as reviews~\citep{zheng2017joint}, tweets~\citep{gong2016hashtag} and news~\citep{cao2017online}, pre-trained word embeddings or pre-trained sentence encoders become indispensable. Some works simply use the average of word embeddings to represent the whole documents~\citep{cenikj2020boosting,brochier2019representation}. Other works focus on the design of task-specific frameworks, where the input word embeddings will be fed into a complex document encoder to generate the document representation~\citep{nguyen2017personalized,xu2016tag,song2016multi,zhang2017joint,tan2016neural}. 
Similarly, the pre-training mechanism is widely used in image feature extraction for recommender systems~\citep{chu2017hybrid,he2016ups,he2016vbpr}.

\subsection{Knowledge Graph-based Recommendation}
Knowledge graph-based recommendation introduces knowledge graphs (KGs) as side-information to better characterize users and items. A KG is a structured graph containing fruitful facts and connections between users, items and other related entities. Amounts of side-information, such as the user profiles, the attributes of items, and relations between cross-domain items, can be integrated into KGs. Hence, KGs can help recommender systems to capture essential knowledge and provide explanations for the recommendation results. 

Various KGs have been used in different works. For instance, some works construct knowledge graphs with items and their related attributes~\citep{zhang2016collaborative,huang2018improving,wang2018dkn}. Some other works add users to build user-item graphs, which contain information including the item attributes, user behaviours (e.g., purchase, click), and user profiles~\citep{wang2018shine,dadoun2019location,cao2019unifying}. With the informative heterogeneous user-item graphs, the potential relations between users and items can be modeled directly.

In order to exploit KGs, one line of KG-based methods seeks to encode the KG into low-dimensional pre-trained embeddings with the knowledge graph embedding (KGE) methods, such as TransE~\citep{bordes2013translating}, TransR~\citep{lin2015transr} and TransH~\citep{wang2014knowledge}. Then as stated in Equation~\ref{eq:aggregate}, the knowledge graph embeddings are aggregated with user/item features obtained from interaction data. 
Experimental results show that KGs are powerful information resource and can improve the performance of recommendation significantly~\citep{qin2020survey}.
\vspace{0.5em}

\subsection{Social Recommendation}
Social recommendation is a type of recommendation methods that utilize online social relations as an additional input~\citep{tang2013social}. Different from KGs which integrate various information about users and items, social graphs focus on modeling the social relation between users. Homophily theory indicates that the preference of a user is similar to or influenced by their socially connected friends~\citep{mcpherson2001birds}. 
Similar to the KG-based recommendation, many social recommender systems seek to integrate the pre-trained social network embeddings, which indicates the degree that a user is influenced by his/her friends~\citep{guo2018exploiting,wen2018network,sathish2019graph,chen2019n2vscdnnr,chen2019co,zhang2018matrix}. 
\vspace{0.5em}

\subsection{Summary}
Feature-based models pre-process side-information with various pre-trained models to obtain the embedding of users or items, which are then integrated into the recommender systems. By utilizing side-information, feature-based approaches are able to construct expressive representations for users and items, and can achieve significant improvement for recommendation. In addition to side-information, exploiting large-scale interaction data is also crucial to recommender systems.
Therefore, recently some efforts have been made to pre-train models with user-item interaction records, which are introduced in the following section.


\section{Fine-tuning Models}
\label{sec:fine-model}
The fine-tuning models for recommendation first pre-train the parameters with large-scale interaction data. The models are then transferred to downstream tasks by simply fine-tuning the pre-trained parameters. The fine-tuning paradigm has shown the effectiveness in other areas, such as natural language processing~\citep{devlin2019bert,joshi2020spanbert}. According to the model architecture, we can categorize existing works in recommender systems into two classes: shallow neural networks~\citep{hu2018conet,ni2018perceive} and deep residual neural networks. Existing deep residual neural networks for recommendation can be further divided into BERT-based models~\citep{sun2019bert4rec,chen2019bert4sessrec,yang2019pre} and parameter-efficient pre-trained convolutional neural networks~\citep{yuan2020parameter}.
\vspace{0.5em}

\subsection{Shallow Neural Networks}
Early works attempt to achieve knowledge transfer with shallow neural networks as base models, such as shallow MLP, recurrent neural networks.
\citet{hu2018conet} attempt to improve recommendation via cross-domain knowledge sharing. They conduct a baseline experiment where an MLP together with user and items embeddings are pre-trained on the source domain. The user embeddings are then transferred to the target domain as warm-up. Experimental results show that this simple method cannot achieve obvious improvement in recommendation performance. The results demonstrate that the model architecture and pre-training tasks need to be carefully designed to achieve effective knowledge transfer in fine-tuning models.
Therefore, many efforts have been devoted to investigating efficient pre-training tasks and transferrable model architectures.



\citet{ni2018perceive} propose the DUPN model which is able to learn universal user representations across multiple recommendation tasks. DUPN takes interaction sequence as inputs, and then applies LSTM and an attention layer to obtain the user representations. DUPN is pre-trained by multiple tasks objectives, including click-through rate prediction, shop preference prediction, price preference prediction. Experimental results show that DUPN can achieve not only improvement on the tasks used for pre-training, but also faster convergence and promising results in related new tasks. Though the user representations learned with DUPN are powerful, DUPN requires many extra information sources like user profiles to facilitate different pre-training tasks.

\subsection{BERT-based Models}
In order to capture the dynamic user preference, many researchers attempt to exploit user chronological interaction sequence, termed as session-based recommendation. Similar to natural language processing (NLP) that targets on word sequence, session-based recommendation investigates item sequence and aims to take sequential information into account.
Inspired by the rapid progress in pre-trained language models in NLP ~\citep{devlin2019bert,liu2019roberta,joshi2020spanbert}, many efforts have been devoted to capturing information from the user behaviour sequence with pre-trained models, especially BERT-based models. In this section, we introduce the pre-trained BERT-based models for recommendation, including the widely used masked item prediction task, the architecture of BERT and advanced BERT-based models for recommendation.

\subsubsection{Masked Item Prediction}
Similar to the masked language modeling task in NLP, the masked item prediction task (MIP) is proposed and widely applied in many recommender systems. In the MIP task, given the interaction sequence, some of the items are randomly masked. The models are required to reconstruct the masked item. Formally, we denote the interaction sequence in chronological order for user $u$ as $\mathcal{S}_u = \{v_1, v_2, ..., v_N\}$, where $v_i$ is the item that $u$ interacted with at time step $i$. For pre-training stage, some input items in the interaction sequence are randomly masked with special token \texttt{[MASK]}. Then the models are asked to predict the masked items. For example:
\begin{align*}
\small
\text{Inputs}&:\quad\{v_1, v_2, v_3, v_4, v_5\} \rightarrow \{v_1, v_2, \texttt{[MASK]}_1, v_4, \texttt{[MASK]}_2\} \\
\text{Labels}&:\quad\ \texttt{[MASK]}_1 = v_3,\qquad \texttt{[MASK]}_2 = v_5
\end{align*}
And the objective of this task is the negative likelihood of the masked targets. Unlike the left-to-right next item prediction task that is used in many session-based recommender systems~\citep{hidasi2018recurrent,wu2017recurrent,yu2016dynamic}, MIP enables the models to learn representations of user behavior sequences from the whole context. Moreover, the MIP task can overcome the limitation that a rigidly ordered sequence is not always practical in user behaviours~\citep{sun2019bert4rec}. Therefore, models pre-trained with MIP task can achieve promising results.

\begin{figure}
    \centering
    \includegraphics{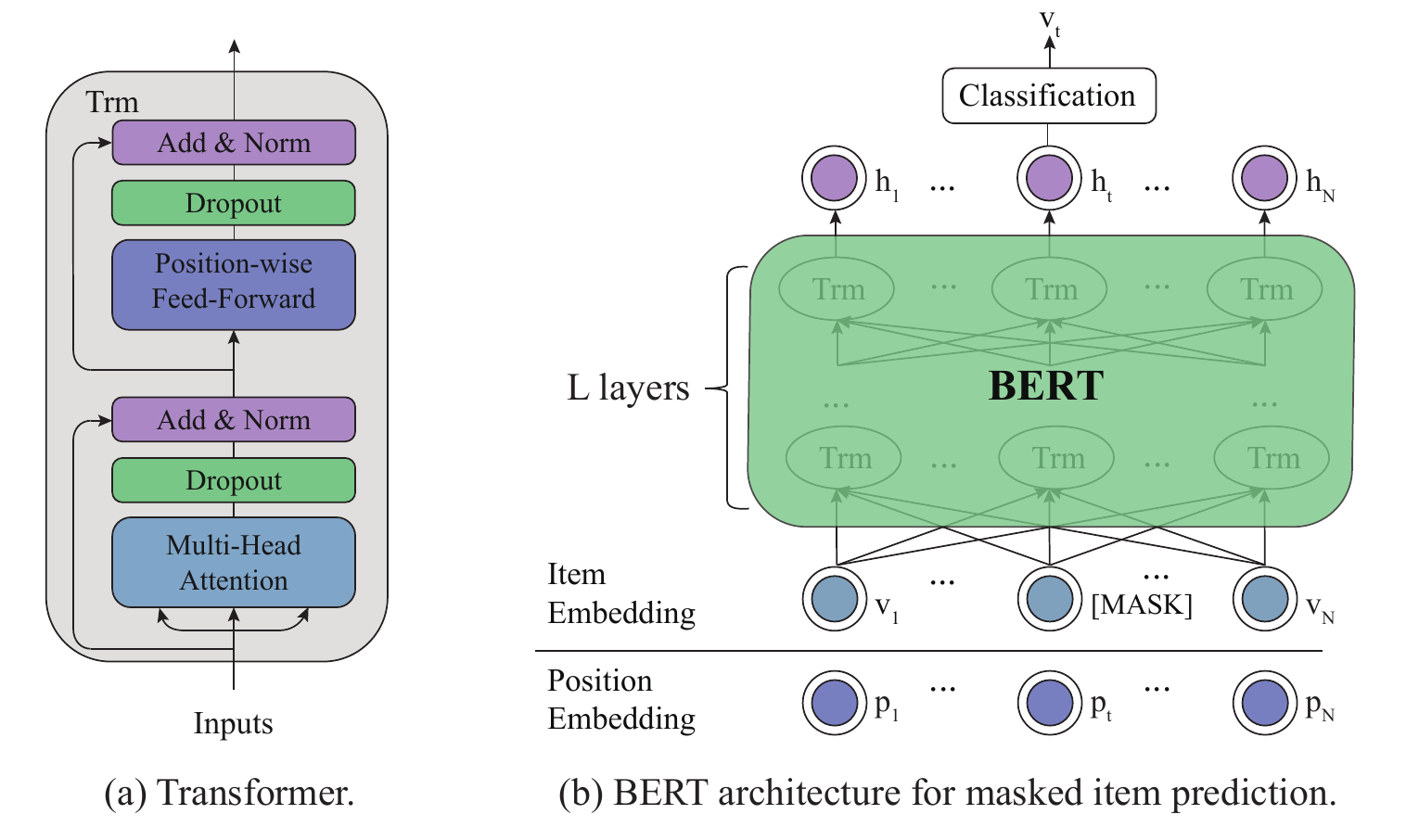}
    \caption{(a) is the framework of Transformer. (b) is the overall architecture of BERT for masked item prediction task.}
    \label{fig:bert4rec}
    \vspace{-1.1em}
\end{figure}
\subsubsection{BERT for Recommender System}
Inspired by the success of BERT~\citep{devlin2019bert} in text understanding, many researchers adopt BERT for recommendation. In this section, we will introduce the architecture of BERT and how to utilize BERT for recommender systems. For convenience, we denote the BERT model pre-trained with the MIP task as BERT4RS. As shown in Figure~\ref{fig:bert4rec}, BERT is based on multi-layer bidirectional Transformer~\citep{vaswani2017attention}. Transformer contains two sub-layers: multi-head attention sub-layer and point-wise feed-forward network.
\begin{itemize}
    \item \textbf{Multi-head Attention.} Attention mechanism has been used successfully in various sequence modeling tasks, which enables models to focus on important information. The attention function takes queries $\mathbf{Q} \in \mathbb{R}^{l_Q \times d_k}$, keys $\mathbf{K} \in \mathbb{R}^{l_K \times d_k}$, and values $\mathbf{V} \in \mathbb{R}^{l_K \times d_v}$ as inputs and computes the outputs as follows, where $d_k$ and $d_v$ are the dimensions, $l_Q$ and $l_K$ are the length of sequences.
    \begin{equation}
        \text{Attention}(\mathbf{Q}, \mathbf{K}, \mathbf{V}) = \text{softmax}(\frac{\mathbf{Q}\mathbf{K}^T}{\sqrt{d_k}})\mathbf{V}
    \end{equation}
    And self-attention is a special attention function aiming to learn the representation for a single sequence, in which the input sequence performs as the queries, keys and values. Instead of performing a single attention function, Transformer employs the multi-head self-attention function, which allows the model to jointly attend to information from different vector sub-spaces. Specifically, this mechanism first linearly projects the input sequence into $h$ sub-spaces and then produces the output representation with $h$ attention functions.
    \begin{align*}
        \text{MultiHead}(\mathbf{H}) &= \text{Concat}(\text{head}_1, \text{head}_2, ..., \text{head}_h)\mathbf{W}^O \\
        \text{head}_i &= \text{Attention}(\mathbf{H}\mathbf{W}_i^Q, \mathbf{H}\mathbf{W}_i^K, \mathbf{H}\mathbf{W}_i^V)
    \end{align*}
    where $\mathbf{H} \in \mathbb{R}^{l \times d_o}$ is the input sequence, $\mathbf{W}_i^Q \in \mathbb{R}^{d_o \times d_k}$, $\mathbf{W}_i^K \in \mathbb{R}^{d_o \times d_k}$, $\mathbf{W}_i^V \in \mathbb{R}^{d_o \times d_v}$, and $\mathbf{W}^O \in \mathbb{R}^{hd_v \times d_o}$ are learnable parameter matrices.
    \item \textbf{Point-wise Feed-Forward Network.} The multi-head attention function enables the model to integrate information from different positions with linear combinations. Then the point-wise feed-forward network endows the model non-linearity. In this sub-layer, a fully connected feed-forward network is applied to each position separately and identically. 
    \begin{equation}
        \text{FFN}(\mathbf{x}) = \text{ReLU}(\mathbf{x}\mathbf{W}_1 + \mathbf{b}_1)\mathbf{W_2} + \mathbf{b}_2
    \end{equation}
    The sub-layer consists of two linear transformations and a ReLU activation. It should be noted that though the transformations are the same across different positions, the parameters are different for different layers.
\end{itemize}

Following each sub-layer, a residual connection~\citep{he2016deep} and a layer normalization operation~\citep{ba2016layer} are employed for stabilizing and accelerating
the network training. As shown in Figure~\ref{fig:bert4rec}b, after $L$ layers of Transformer, the final hidden states of masked tokens are fed into a feedforward network to get the output distribution over target items.

BERT4RS is effective in modeling user preference from the historical behaviours. \citet{sun2019bert4rec} train a two-layer BERT with the MIP task, which achieves the state-of-the-art performance on the session-based next item recommendation task. They observe that both the BERT architecture and the MIP task can significantly improve the performance, and stacking multiple Transformer layers can further boost the performance on large-scale datasets, which provides fundamental support for the following models.

\citet{chen2019bert4sessrec} propose to fine-tune BERT4RS with a content-based click through prediction task. Specifically, the user representation $\mathbf{u}$ is produced with pre-trained BERT from the historical behaviour sequence, and the item representation $\mathbf{v}$ is extracted from its content. Then the preference score is calculated with Equation~\ref{eq:score}, where $f(\cdot)$ is a MLP layer.

Some downstream recommendation tasks, such as next basket recommendation~\citep{rendle2010factorizing,yu2016dynamic} and list-wise recommendation~\citep{shi2010list,zhao2017deep}, require the model to capture the relations between item sequences. Therefore, in addition to MIP, researchers propose some sequence-level pre-training tasks to pre-train the model. \citet{yang2019pre} adopts the BERT4RS for next basket recommendation task, which is pre-trained with MIP and next basket prediction (NBP) tasks. In real-world scenarios, a user usually buys or browses a series of items (a basket) at a time. Given two baskets, NBP requires the model to predict whether the two baskets are adjacent in the purchase records. For example:
\begin{align*}
\small
    \text{Inputs}&:\quad \ \{\texttt{[CLS]}, \quad v_1^1, \, \quad \ v_2^1, \ \; \quad v_3^1, \ \quad \texttt{[SEP]}, \ \quad v_1^2, \ \quad v_2^2, \ \; \quad v_3^2, \quad \ \ \texttt{[SEP]}\} \\
    \rightarrow&\qquad \{\texttt{[CLS]}, \quad v_1^1, \texttt{[MASK]}_1, v_3^1, \quad \ \texttt{[SEP]}, \ \quad v_1^2, \ \quad v_2^2, \texttt{[MASK]}_2, \texttt{[SEP]}\} \\
    \text{MIP Labels}&: \quad \texttt{[MASK]}_1 = v_2^1, \qquad \texttt{[MASK]}_2 = v_3^2 \\
    \text{NBP Label}&: \quad \texttt{IsNext/NotNext}
\end{align*}
where $v_i^1$ and $v_i^2$ are items from different baskets, \texttt{[CLS]} and \texttt{[SEP]} are special tokens. The final hidden state of \texttt{[CLS]} is used to predict the NBP label. 


\subsection{Parameter-Efficient Pre-trained Model}
The pre-training mechanism can enable models to capture user preference from behaviour history via self-supervised learning. Experimental results show that it can achieve significant improvement for recommendation. However, fine-tuning models separately for different tasks is computationally expensive and memory intensive~\citep{stickland2019bert,mudrakarta2018k}, especially for resource-limited devices. 

\begin{figure}[t]
    \centering
    \includegraphics[width=\linewidth]{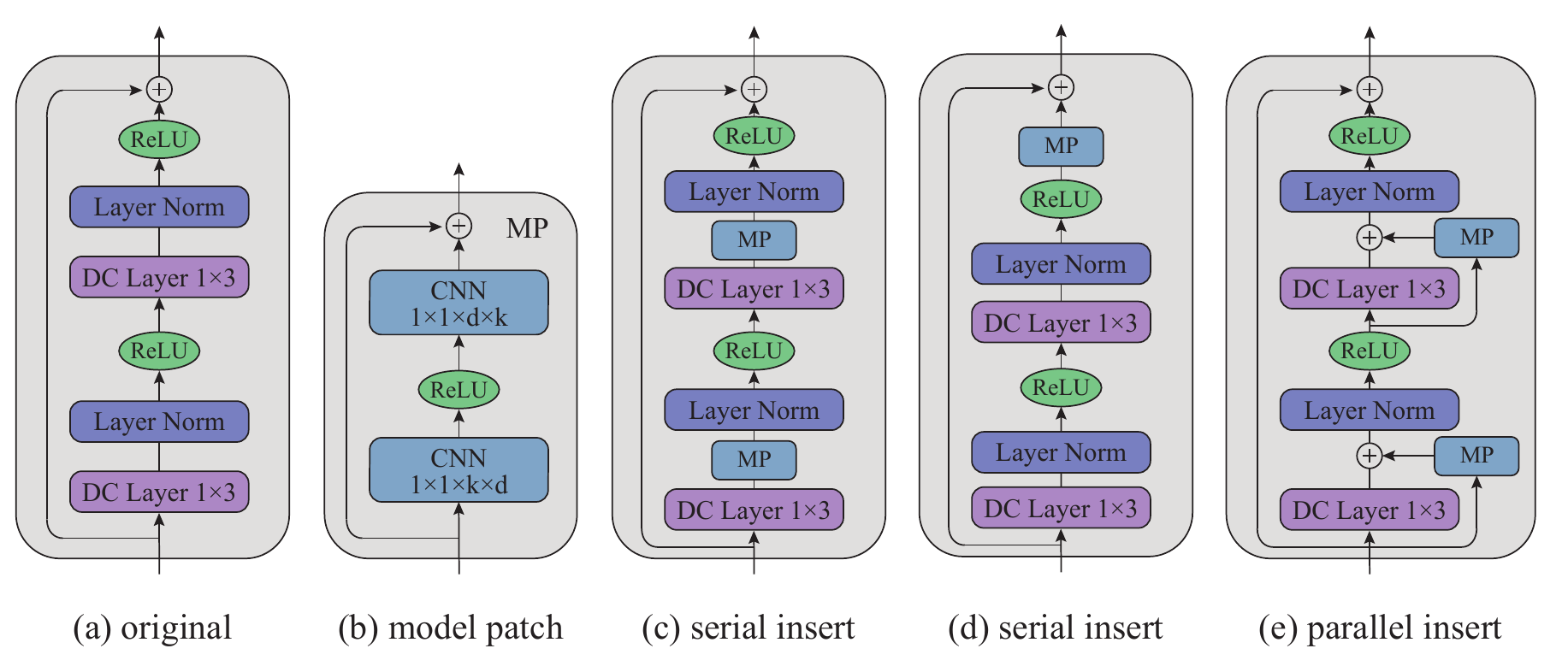}
    \caption{The model architecture of Peterrec~\citep{yuan2020parameter}. (a) is the original residual block of the pre-trained model. (b) is the model patch(MP) block. (c)(d)(e) are different approaches to insert MP blocks for fine-tuning. $\bigoplus$ indicates element-wise addition operation. $1\times3$ and $1\times1\times d \times k$ refer to the kernel size of convolutional layers.}
    \label{fig:peterrec}
\end{figure}

To address this issue, \citet{yuan2020parameter} propose the Peterrec, which utilizes a grafting neural network in fine-tuning, termed as model patch (MP). By inserting MPs in the pre-trained models, the fine-tuning networks can keep all pre-trained parameters unchanged. As shown in Figure~\ref{fig:peterrec}, Peterrec is a stack of dilated convolutional (DC) layers, and there is a residual connection between every two DC layers. Similar to other pre-trained models, Peterrec employs the MIP task to pre-train. During the fine-tuning stage, the MP blocks, which is a simple residual two-layer convolutional network, are inserted around the original DC layers. The pre-trained parameters are shared across different tasks. Only parameters of the MP blocks will be fine-tuned. To accelerate fine-tuning and minimize the number of parameters, the MP blocks are designed in a bottleneck architecture. Specifically, the first convolutional layer projects the $k$ dimensional channels into $d$ ($d \ll k$) dimensional vectors, and the second layer projects it back to its original dimension. Thus, the number of inserted parameters can be less than $10\%$ parameters of original pre-trained models.
Empirical results show that the pre-trained Peterrec are useful for various downstream tasks, including user profile prediction and top-k recommendation. Moreover, it can achieve promising results even when the user is cold in the new target domain, which proves the effectiveness of pre-trained models in knowledge transfer for recommendation.

\subsection{Summary}
The pre-training mechanism works well in many recommendation tasks. Early works explore pre-training with shallow neural networks. Inspired by the success of pre-trained language models in NLP, the deep residual models pre-trained with MIP task are widely used in many different recommender systems. Although it has been proven effective to employ pre-trained models for recommendation, there are many open challenges to be addressed, which will be discussed in Section~\ref{sec:outlook}.

\section{Experiment of Recommender System with Pre-training}
\label{sec:experiments}

In this section, we conduct experiments to verify the benefits of pre-training to recommender systems. We take next-item recommender systems as example, and investigate the potential of pre-trained recommender systems in cold start problem and cross-domain knowledge transfer.

\subsection{Dataset}

We evaluate the models on MovieLens\footnote{\url{https://grouplens.org/datasets/movielens/}}, a representative and popular real-world benchmark in recommendation field. We choose the well-established lightweight version MovieLens 1m (\textbf{ML-1m}). We follow \cite{tang2018personalized} for data preprocessing. Users and items with too few ratings ($<5$) are filtered out. We leave the last $5$ items of each sequence to form the validation set ($2$ items for each user) and the test set ($3$ items for each user). To investigate the effectiveness of pre-trained recommender systems in cross-domain context, we divide all the movies in ML-1m into two domains according to movie genres, and obtain interaction sequences of the same users on two domains. The scale of the target domain (\textbf{ML-1m-tgt}) is much smaller than the source domain (\textbf{ML-1m-src}), with shorter sequence, thus it naturally forms a user cold-start situation for movie items in target domain. 

Apart from user-item interactions, we also use the meta information from IMDB\footnote{\url{https://www.kaggle.com/carolzhangdc/imdb-5000-movie-dataset}} (\textbf{IMDB 5000 Movie Dataset}) to provide side-information for the items. For each movie, we inquire the director and three main actors. The statistics of the processed datasets are summarized in Table~\ref{table1}.

\begin{table*}[h]
\renewcommand\arraystretch{1.25}
\begin{center}
\setlength{\tabcolsep}{1.5mm}{
\begin{tabular}{c|ccccc}

\toprule
Dataset & \#User & \#Item & \#ItMeta & \#InA & AvgLen\\
\midrule
ML-1m & 6,040 & 3,416 & 2,025 & 999k & 165.5 \\
\midrule 
ML-1m-src & 6,040 & 2,349 & 1,415 & 725k & 120.1\\
\midrule 
ML-1m-tgt & 6,040 & 1,067 & 610 & 274k & 45.4\\
\bottomrule 
\end{tabular}}
\end{center}
\caption{Statistics of datasets. \#User, \#Item, \#ItMeta, and \#InA indicate the number of users, items, items with meta information, and interactions, respectively. AvgLen refers to the average length of input sequences.}
\label{table1}
\end{table*}

\subsection{Task Settings and Baselines}
We regard user ratings as interaction sequence based on timestamps, and adopt next item recommendation task to evaluate sequence models. For each user sequence, we generate negative samples in which items are not seen in training data. Since movie rating behavior is not strictly sequential, we treat the $3$ items in each test sequence as equivalent. We adopt Normalized Discounted Cumulative Gain (NDCG@K) and Recall@K as evaluation metrics. For pre-training, models are provided with ML-1m-src as pre-training data and IMDB 5000 as side-information for items. Models are fine-tuned and evaluated on ML-1m-tgt. 

For baseline methods, we choose two representative models. \textbf{Caser}~\citep{tang2018personalized} is a shallow recommendation model which combines vertical and horizontal CNNs to extract user and item representations.  \textbf{BERT4Rec}~\citep{sun2019bert4rec} utilizes BERT as architecture backbone for session-based recommendation. To investigate the potential of pre-training in deep recommendation models, we increase the number of layers in BERT4Rec. 

\subsection{Model Design Choices}
We investigate the effect of different design choices to pre-trained recommendation systems, including knowledge transfer approach, pre-training tasks. We also explore the potential of pre-trained recommendation systems in integrating external side-information.

\textbf{Knowledge Transfer Approach.} Our main purpose is to verify the effectiveness of pre-training on recommender systems. Therefore, we first investigate pre-training session-based recommendation models on ML-1m-src before fine-tuning on ML-1m-tgt. We hope that the knowledge learned from pre-training can be transferred between domains. During fine-tuning, we investigate two approaches to knowledge transfer: (1) \textit{Shallow Transfer.} We can transfer and fine-tune only the input embedding layer (i.e., user embeddings and external knowledge embeddings) and leave other layers randomly initialized and learned from scratch in the target domain. (2) \textit{Deep Transfer.} We also explore transferring and fine-tuning the whole model on the target domain. In this way, user-item interaction knowledge is also expected to be transferred\footnote{Note that in both knowledge transfer approaches, item embeddings are randomly initialized on the target domain, since items are not shared between domains.}.

\textbf{Pre-training Tasks.} The design of pre-training tasks is crucial for pre-trained models to capture general knowledge from large-scale data. We compare two widely used pre-training tasks in recommender systems: (1) \textit{Next Item Prediction} (NIP). Next item prediction recurrently predicts the next item in a left-to-right fashion. (2) \textit{Masked Item Prediction} (MIP). Masked item prediction randomly masks the items and predicts the masked item according to bidirectional context. 

\textbf{External Knowledge.} External knowledge is shown to be effective in handling data sparsity problem in recommender systems. We investigate whether external knowledge can be combined with general knowledge learned from pre-training to achieve better results. Specifically, we constrain the item features by concatenating the external knowledge embeddings (i.e., director embeddings and actor embeddings) with the 
item embeddings to obtain external knowledge enhanced item embeddings.

\subsection{Implementation Details}

We find the optimal settings of hyperparameters via grid search on the validation set.

For Caser model, we use implementation provided by the authors\footnote{\url{https://github.com/graytowne/caser_pytorch}}. Batch size is searched from  $\{128, 256, 512, 1024\}$ and tuned as $512$. Hidden dimension size is searched from $\{30, 50, 100, 150\}$ and tuned as $100$. We employ Adam to optimize the model, and we set the learning rate as $1e\text{-}3$, weight decay as $1e\text{-}6$.
For masked item prediction, we set the mask probability as $0.2$. Following \citet{devlin2019bert}, we do not always replace the masked item with \texttt{[MASK]} token. If an item is chosen, we replace it with (1) \texttt{[MASK]} $80\%$ of the time (2) a random item $10\%$ of the time (3) the unchanged item $10\%$ of the time.

For BERT4Rec~\citep{sun2019bert4rec} model, we choose a PyTorch implementation\footnote{\url{https://github.com/jaywonchung/BERT4Rec-VAE-Pytorch}}. We keep most of the original hyper-parameter and initialization strategy settings. Batch size is searched from $\{128, 256, 512\}$ and tuned as $256$. Hidden dimension size is searched from $\{64, 128, 256\}$ and tuned as $128$. We employ Adam to optimize the model, and we set the learning rate as $1e\text{-}3$. We add user embedding and set the user embedding size as $32$. We change the layer of BERT blocks to $6$. For masked item prediction, we set the mask probability as $0.15$ and use the same masking strategy as in Caser.

When incorporating external knowledge, the size of director and main actors embedding is $25$ for Caser and $16$ for BERT4Rec. The number of negative samples if set to $100$. In evaluation, negative samples size is $100$ for both models.

\subsection{Experiment Results}
The experiment results on ML-1m-tgt are reported in Table~\ref{tab:main results}, from which we have the following observations\footnote{The observations are supported by statistical significance test on the experiment results with $p < 0.05$.}:


\begin{table*}[t]
\centering
\small
\resizebox{\textwidth}{!}{
\begin{tabular}{ll|cc|c|rrrrr}
\toprule
\multirow{2}{*}{Model}          & \multirow{2}{*}{Transfer} & \multicolumn{2}{c|}{Task} & \multirow{2}{*}{With Meta} &  \multicolumn{5}{c}{Test Result}  \\
& \multicolumn{1}{c|}{}  & NIP & MIP & & \tiny{NDCG@1}  &  \tiny{NDCG@5}  & \tiny{NDCG@10}  & \small{Recall@5}  & \small{Recall@10}       
\\ \midrule
\multirow{10}{*}{\shortstack{Caser}} & 
\multirow{2}{*}{\shortstack{None}} &
 & & & 3.27 & 6.09 & 7.47 & 8.83  & 13.14   \\
  & & & & \checkmark &  3.17  & 6.14  & 7.53  & 9.00  & 13.32  \\

\cmidrule(lr{1em}){2-10}

 & \multirow{4}{*}{\shortstack{Shallow}} &
 \multirow{2}{*}{\checkmark} & & & \textbf{3.33}  & 6.24 & \textbf{7.67} & 9.06  & 13.53  \\
  & & & & \checkmark & 3.27  & \textbf{6.25} & 7.62 & 9.21 & 13.49  \\
   \cmidrule(lr{1em}){3-10}
 & & & \multirow{2}{*}{\checkmark} & &  3.18  & 6.25  & 7.62  & \textbf{9.26}  & 13.54  \\
 & & & & \checkmark & 3.13  & 6.09 & 7.51  & 8.88 & 13.28  \\
 
 \cmidrule(lr{1em}){2-10}
 
 & \multirow{4}{*}{\shortstack{Deep}} &
  \multirow{2}{*}{\checkmark} & & & 2.87  & 5.82 & 7.34 & 8.75 & 13.47 \\
  & & & & \checkmark & 3.21 & 6.08 & 7.59 & 8.89 & \textbf{13.59} \\
   \cmidrule(lr{1em}){3-10}
 & & & \multirow{2}{*}{\checkmark} & &  2.73  & 5.76 & 7.18 & 8.78 & 13.18 \\
 & & & & \checkmark & 3.13  & 6.06 & 7.48 & 8.90 & 13.32 \\
 
 \midrule
 
 \multirow{10}{*}{\shortstack{BERT4Rec}} & 
\multirow{2}{*}{\shortstack{None}} &
 & & & 30.07  & 32.05  & 39.44  & 34.99  & 51.35  \\
  & & & & \checkmark &  30.18  & 32.75 & 39.94  & 35.83  & 51.72\\

\cmidrule(lr{1em}){2-10}

 & \multirow{4}{*}{\shortstack{Shallow}} &
 \multirow{2}{*}{\checkmark} & & & 30.92  & 33.93  & 41.37  & 37.48  & 53.92  \\ 
  & & & & \checkmark & 32.98  & 35.58  & 42.85  & 38.98  & \textbf{55.02}  \\
   \cmidrule(lr{1em}){3-10}
 & & & \multirow{2}{*}{\checkmark} & &  31.03  & 34.00 & 41.39 & 37.40  & 53.73  \\
 & & & & \checkmark & 32.61  & 35.19  & 42.36  & 38.51  & 54.35  \\
 
 \cmidrule(lr{1em}){2-10}
 
 & \multirow{4}{*}{\shortstack{Deep}} &
  \multirow{2}{*}{\checkmark} & & & 30.57  & 33.69 & 41.21  & 37.28 & 53.90  \\
  & & & & \checkmark & 32.96  & 35.31  & 42.71  & 38.55  & 54.91  \\
   \cmidrule(lr{1em}){3-10}
 & & & \multirow{2}{*}{\checkmark} & &  32.47  & 35.33 & 42.60 & 38.69 & 54.73 \\
 & & & & \checkmark & \textbf{33.54} & \textbf{35.84}  & \textbf{42.95}  & \textbf{39.04}  & 54.73  \\

\bottomrule
\end{tabular}
}
\caption{Performance (\%) comparison of different settings on next-item prediction of ML-1m-tgt \protect\footnotemark[7].}
\label{tab:main results}
\end{table*}
\vspace{-0.5em}

\newcommand{\tabincell}[2]{\begin{tabular}{@{}#1@{}}#2\end{tabular}}

(1) Pre-training boosts the recommendation performance for both models. However, the effects of knowledge transfer approaches are correlated with model capacity. Specifically, for the shallow Caser model, shallow transfer (i.e., only transfer the emebdding layer) achieves larger improvements with deep transfer (i.e., transfer the whole model). In contrast, deep transfer achieves much better performance for deep BERT4Rec model. We hypothesize that general knowledge about user-item interactions can be better captured by high-capacity models with pre-training, leading to better performance in transferring whole models.

(2) For pre-training tasks, masked item prediction achieves better performance than next item prediction for BERT-based models, which is consistent with the results reported by~\citet{sun2019bert4rec}. One possible reason is that movie rating behavior does not strictly follows chronological order (i.e., the chronological order of two items are likely to be swapped). Therefore, by integrating bidirectional information, masked item prediction can learn better representations during pre-training. However, next item prediction seems better than masked item prediction for Caser. It is probably because masked item prediction creates gap between pre-training and fine-tuning (e.g., prediction during fine-tuning is based on unidirectional context), which cannot be easily overcome by shallow models. 

(3) Incorporating external knowledge improves the performance of pre-trained BERT4Rec model. We note that for pre-trained Caser, the effect of external knowledge is not significant. We speculate that the limited capacity of simple models hinders the integration of external knowledge. Besides, simple concatenation cannot well integrate external knowledge with general knowledge learned from pre-training either. More advanced methods can be developed to inject external knowledge into pre-trained recommendation models.





\footnotetext[7]{Omit Recall@1 which is identical to NDCG@1.}

In conclusion, recommender systems could benefit from pre-training, which effectively transfer knowledge between domains and tasks, and has potential in solving problems including cold start. 

\section{Open Challenges and Future Directions}
\label{sec:outlook}
Although the pre-trained models have shown their power in recommender systems, the challenges still exists. In this section, we suggest five open challenges and future directions of pre-trained models in recommendation, where (1)(2)(3) are discussion about how to better utilize pre-trained models in various recommendation scenarios and (4)(5) mainly focus on how to improve the pre-training mechanism to better serve recommender systems.
\begin{itemize}
    \item[(1)] \textbf{Cold Start.} Collaborative filtering recommender systems rely heavily on user historical behaviour data, and can suffer from the cold start problem. To alleviate the problem, some approaches~\citep{ma2011recommender,manotumruksa2016regularising,yu2018adaptive} use side-information, such as user profile and item attributes, to enrich user/item representations. Besides, there are some efforts utilizing more efficient learning mechanism to alleviate the heavy data reliance, such as few-shot learning~\citep{lee2019melu,li2019zero}.
    
    Pre-trained language models can significantly improve few-shot performance~\citep{brown2020language} in NLP tasks. Similarly, in terms of recommendation, pre-trained models can be applied for cold start problem by learning transferable representations of the shared information between large-scale general domain and sparse target domain. For example, if a user is cold in the target domain, it is useful to transfer his/her representation pre-trained in the general domain; if an item is cold, its representation can be estimated by leveraging the pre-trained representations of external information.
    Peterrec~\citep{yuan2020parameter} is a good exploration which achieve user cold start with pre-trained models.
    
    \item[(2)] \textbf{Knowledge Enhanced Pre-training.} Knowledge graphs can provide rich domain knowledge, world knowledge and commonsense knowledge for recommendation. Therefore, by incorporating KGs into recommendation, user preference and relations between items can be captured more accurately. Many KG-based approaches are proposed recently, and have achieve promising results~\citep{qin2020survey,zhang2016collaborative,wang2018dkn,tang2019akupm}. However, few works consider directly injecting external structured knowledge into pre-trained models for recommendation.
    
    In fact, many knowledge enhanced pre-trained language models~\citep{zhang2019ernie,liu2020k,wang2020k} have shown that fusing the structured knowledge into pre-trained models can significantly boost the performance of original models.
    Knowledge information can help models better characterize users and items, and thus can improve the performance of recommendation.
    
    \item[(3)] \textbf{Social Relation Enhanced Pre-training.} Social relations provide a possible perspective for personalized recommendation. Users who are connected are more likely to share similar preferences.
    Pre-trained models are proficient in capturing user interest from their historical interaction records. Therefore, the social relations between users can be viewed as meta-relations between user-item interaction sequences, i.e., the interaction sequences of closely connected users are encouraged to share similar representations. Based on this, sequence-level pre-training tasks can be proposed to help models to generate more expressive user/item representations.
    
    Another possible direction is to employ social relation enhanced pre-trained models to solve user cold start problem. Social relations can provide clues for the user interest. However, it is still challenging to make full use of the rich information contained in the neighboring users in social relation graphs during pre-training. 
    
    \item[(4)] \textbf{Pre-training Tasks.} Currently, all the deep fine-tuning approaches rely on the MIP task to pre-train the model. And these works focus on extracting user interest from their historical sequential records. However, limited by the computing ability and memory of GPUs, only the most recent interaction records that represent recent user preference, can be utilized by recommendation models. Besides, MIP can only utilize sequential data, while rich heterogeneous information is usually available in many real-world scenarios. Therefore, designing new self-supervised pre-training tasks is important to make full use of the large-scale heterogeneous data for recommendation.
    
    \item[(5)] \textbf{Model Architecture and Model Compression.} The pre-trained models are effective in various recommendation tasks. However, their high computation complexity makes it hard to deploy them in real-world scenarios. To address the problem, it would be helpful to perform model compression~\citep{gordon2020compressing,mccarley2019pruning} or improve the model architecture. Besides, fine-tuning separately for each down-stream task is quite time-consuming and memory intensive. The model patch~\citep{yuan2020parameter} is a good attempt to reduce memory cost. However, it is still an urgent need to achieve fast and effective knowledge transfer from pre-trained models to multiple down-stream tasks.
    
\end{itemize}

\section{Conclusion}
In this paper, we investigate the pre-trained models for recommendation and summarize the efforts devoted to this direction. We conduct comprehensive overview of two types of pre-trained models for recommendation, including feature-based models and fine-tuning models. Then we conduct experiments to show the benefits of pre-training for recommender systems. Finally, open challenges and future direction are discussed, hoping to promote the progress of this domain.

\section*{Author Contributions}

ZZ, CX, YY and RX contributed to the paper structure design and writing. CX collected the related papers. ZZ conducted the experiments. ZL, FL, LL and MS provided valuable suggestions and helped in revising the paper.

\section*{Funding}
This work is supported by the National Key Research and Development Program of China (No. 2018YFB1004503).




\bibliographystyle{frontiersinSCNS_ENG_HUMS} 
\bibliography{frontiers}

\begin{thebibliography}{73}
\providecommand{\natexlab}[1]{#1}
\expandafter\ifx\csname urlstyle\endcsname\relax
  \providecommand{\doi}[1]{doi:\discretionary{}{}{}#1}\else
  \providecommand{\doi}{doi:\discretionary{}{}{}\begingroup
  \urlstyle{rm}\Url}\fi
\providecommand{\selectlanguage}[1]{\relax}
\providecommand{\bibAnnoteFile}[1]{%
  \IfFileExists{#1}{\begin{quotation}\noindent\textsc{Key:} #1\\
  \textsc{Annotation:}\ \input{#1}\end{quotation}}{}}
\providecommand{\bibAnnote}[2]{%
  \begin{quotation}\noindent\textsc{Key:} #1\\
  \textsc{Annotation:}\ #2\end{quotation}}

\bibitem[{Ba et~al.(2016)Ba, Kiros, and Hinton}]{ba2016layer}
Ba, J.~L., Kiros, J.~R., and Hinton, G.~E. (2016).
\newblock Layer normalization.
\newblock \emph{arXiv preprint arXiv:1607.06450}
\bibAnnoteFile{ba2016layer}

\bibitem[{Bordes et~al.(2013)Bordes, Usunier, Garcia-Duran, Weston, and
  Yakhnenko}]{bordes2013translating}
Bordes, A., Usunier, N., Garcia-Duran, A., Weston, J., and Yakhnenko, O.
  (2013).
\newblock Translating embeddings for modeling multi-relational data.
\newblock In \emph{Proceedings of the NIPS}. 2787--2795
\bibAnnoteFile{bordes2013translating}

\bibitem[{Brochier(2019)}]{brochier2019representation}
Brochier, R. (2019).
\newblock Representation learning for recommender systems with application to
  the scientific literature.
\newblock In \emph{Proceedings of the WWW}. 12--16
\bibAnnoteFile{brochier2019representation}

\bibitem[{Brown et~al.(2020)Brown, Mann, Ryder, Subbiah, Kaplan, Dhariwal
  et~al.}]{brown2020language}
Brown, T.~B., Mann, B., Ryder, N., Subbiah, M., Kaplan, J., Dhariwal, P.,
  et~al. (2020).
\newblock Language models are few-shot learners.
\newblock \emph{arXiv preprint arXiv:2005.14165}
\bibAnnoteFile{brown2020language}

\bibitem[{Cantador et~al.(2015)Cantador, Fern{\'a}ndez-Tob{\'\i}as, Berkovsky,
  and Cremonesi}]{cantador2015cross}
Cantador, I., Fern{\'a}ndez-Tob{\'\i}as, I., Berkovsky, S., and Cremonesi, P.
  (2015).
\newblock Cross-domain recommender systems.
\newblock In \emph{Recommender systems handbook} (Springer), 919--959
\bibAnnoteFile{cantador2015cross}

\bibitem[{Cao et~al.(2017)Cao, Yang, and Liu}]{cao2017online}
Cao, S., Yang, N., and Liu, Z. (2017).
\newblock Online news recommender based on stacked auto-encoder.
\newblock In \emph{Proceedings of the ICIS} (IEEE), 721--726
\bibAnnoteFile{cao2017online}

\bibitem[{Cao et~al.(2019)Cao, Wang, He, Hu, and Chua}]{cao2019unifying}
Cao, Y., Wang, X., He, X., Hu, Z., and Chua, T.-S. (2019).
\newblock Unifying knowledge graph learning and recommendation: Towards a
  better understanding of user preferences.
\newblock In \emph{Proceedings of the WWW}. 151--161
\bibAnnoteFile{cao2019unifying}

\bibitem[{Cenikj and Gievska(2020)}]{cenikj2020boosting}
Cenikj, G. and Gievska, S. (2020).
\newblock Boosting recommender systems with advanced embedding models.
\newblock In \emph{Proceedings of the Web Conference}. 385--389
\bibAnnoteFile{cenikj2020boosting}

\bibitem[{Chen et~al.(2019{\natexlab{a}})Chen, Chen, Huang, Fang, Li, Liu
  et~al.}]{chen2019co}
Chen, J., Chen, W., Huang, J., Fang, J., Li, Z., Liu, A., et~al.
  (2019{\natexlab{a}}).
\newblock Co-purchaser recommendation based on network embedding.
\newblock In \emph{Proceedings of the WISE} (Springer), 197--211
\bibAnnoteFile{chen2019co}

\bibitem[{Chen et~al.(2019{\natexlab{b}})Chen, Wu, Fan, Lin, Zheng, Yu
  et~al.}]{chen2019n2vscdnnr}
Chen, J., Wu, Y., Fan, L., Lin, X., Zheng, H., Yu, S., et~al.
  (2019{\natexlab{b}}).
\newblock N2vscdnnr: A local recommender system based on node2vec and rich
  information network.
\newblock \emph{IEEE Transactions on Computational Social Systems} 6, 456--466
\bibAnnoteFile{chen2019n2vscdnnr}

\bibitem[{Chen et~al.(2019{\natexlab{c}})Chen, Liu, Lei, Li, Zha, and
  Xiong}]{chen2019bert4sessrec}
Chen, X., Liu, D., Lei, C., Li, R., Zha, Z.-J., and Xiong, Z.
  (2019{\natexlab{c}}).
\newblock Bert4sessrec: Content-based video relevance prediction with
  bidirectional encoder representations from transformer.
\newblock In \emph{Proceedings of the MM}. 2597--2601
\bibAnnoteFile{chen2019bert4sessrec}

\bibitem[{Chu and Tsai(2017)}]{chu2017hybrid}
Chu, W.-T. and Tsai, Y.-L. (2017).
\newblock A hybrid recommendation system considering visual information for
  predicting favorite restaurants.
\newblock In \emph{Proceedings of the CIKM}. vol.~20, 1313--1331
\bibAnnoteFile{chu2017hybrid}

\bibitem[{Dadoun et~al.(2019)Dadoun, Troncy, Ratier, and
  Petitti}]{dadoun2019location}
Dadoun, A., Troncy, R., Ratier, O., and Petitti, R. (2019).
\newblock Location embeddings for next trip recommendation.
\newblock In \emph{Proceedings of the WWW}. 896--903
\bibAnnoteFile{dadoun2019location}

\bibitem[{Devlin et~al.(2019)Devlin, Chang, Lee, and
  Toutanova}]{devlin2019bert}
Devlin, J., Chang, M.-W., Lee, K., and Toutanova, K. (2019).
\newblock Bert: Pre-training of deep bidirectional transformers for language
  understanding.
\newblock In \emph{Proceedings of the NAACL}. 4171--4186
\bibAnnoteFile{devlin2019bert}

\bibitem[{Fan et~al.(2018)Fan, Li, and Cheng}]{fan2018deep}
Fan, W., Li, Q., and Cheng, M. (2018).
\newblock Deep modeling of social relations for recommendation.
\newblock In \emph{Proceedings of the AAAI} (AAAI press), 8075--8076
\bibAnnoteFile{fan2018deep}

\bibitem[{Gong and Zhang(2016)}]{gong2016hashtag}
Gong, Y. and Zhang, Q. (2016).
\newblock Hashtag recommendation using attention-based convolutional neural
  network.
\newblock In \emph{Proceedings of the IJCAI}. 2782--2788
\bibAnnoteFile{gong2016hashtag}

\bibitem[{{Gope} and {Jain}(2017)}]{8229786}
{Gope}, J. and {Jain}, S.~K. (2017).
\newblock A survey on solving cold start problem in recommender systems.
\newblock In \emph{Proceedings of the ICCCA}. 133--138
\bibAnnoteFile{8229786}

\bibitem[{Gordon et~al.(2020)Gordon, Duh, and Andrews}]{gordon2020compressing}
Gordon, M.~A., Duh, K., and Andrews, N. (2020).
\newblock Compressing bert: Studying the effects of weight pruning on transfer
  learning.
\newblock \emph{arXiv preprint arXiv:2002.08307}
\bibAnnoteFile{gordon2020compressing}

\bibitem[{Guo et~al.(2018)Guo, Wen, and Wang}]{guo2018exploiting}
Guo, L., Wen, Y.-F., and Wang, X.-H. (2018).
\newblock Exploiting pre-trained network embeddings for recommendations in
  social networks.
\newblock \emph{Journal of Computer Science and Technology} 33, 682--696
\bibAnnoteFile{guo2018exploiting}

\bibitem[{Guo et~al.(2020)Guo, Zhuang, Qin, Zhu, Xie, Xiong
  et~al.}]{guo2020survey}
Guo, Q., Zhuang, F., Qin, C., Zhu, H., Xie, X., Xiong, H., et~al. (2020).
\newblock A survey on knowledge graph-based recommender systems.
\newblock \emph{arXiv preprint arXiv:2003.00911}
\bibAnnoteFile{guo2020survey}

\bibitem[{He et~al.(2016)He, Zhang, Ren, and Sun}]{he2016deep}
He, K., Zhang, X., Ren, S., and Sun, J. (2016).
\newblock Deep residual learning for image recognition.
\newblock In \emph{Proceedings of the CVPR}. 770--778
\bibAnnoteFile{he2016deep}

\bibitem[{He and McAuley(2016{\natexlab{a}})}]{he2016ups}
He, R. and McAuley, J. (2016{\natexlab{a}}).
\newblock Ups and downs: Modeling the visual evolution of fashion trends with
  one-class collaborative filtering.
\newblock In \emph{Proceedings of the WWW}. 507--517
\bibAnnoteFile{he2016ups}

\bibitem[{He and McAuley(2016{\natexlab{b}})}]{he2016vbpr}
He, R. and McAuley, J. (2016{\natexlab{b}}).
\newblock Vbpr: visual bayesian personalized ranking from implicit feedback.
\newblock In \emph{Proceedings of the AAAI}. 144--150
\bibAnnoteFile{he2016vbpr}

\bibitem[{Hidasi and Karatzoglou(2018)}]{hidasi2018recurrent}
Hidasi, B. and Karatzoglou, A. (2018).
\newblock Recurrent neural networks with top-k gains for session-based
  recommendations.
\newblock In \emph{Proceedings of the CIKM}. 843--852
\bibAnnoteFile{hidasi2018recurrent}

\bibitem[{Hu et~al.(2018)Hu, Zhang, and Yang}]{hu2018conet}
Hu, G., Zhang, Y., and Yang, Q. (2018).
\newblock Conet: Collaborative cross networks for cross-domain recommendation.
\newblock In \emph{Proceedings of the CIKM}. 667--676
\bibAnnoteFile{hu2018conet}

\bibitem[{Huang et~al.(2018)Huang, Zhao, Dou, Wen, and
  Chang}]{huang2018improving}
Huang, J., Zhao, W.~X., Dou, H., Wen, J.-R., and Chang, E.~Y. (2018).
\newblock Improving sequential recommendation with knowledge-enhanced memory
  networks.
\newblock In \emph{Proceedings of the SIGIR}. 505--514
\bibAnnoteFile{huang2018improving}

\bibitem[{Joshi et~al.(2020)Joshi, Chen, Liu, Weld, Zettlemoyer, and
  Levy}]{joshi2020spanbert}
Joshi, M., Chen, D., Liu, Y., Weld, D.~S., Zettlemoyer, L., and Levy, O.
  (2020).
\newblock Spanbert: Improving pre-training by representing and predicting
  spans.
\newblock \emph{TACL} 8, 64--77
\bibAnnoteFile{joshi2020spanbert}

\bibitem[{Lee et~al.(2019)Lee, Im, Jang, Cho, and Chung}]{lee2019melu}
Lee, H., Im, J., Jang, S., Cho, H., and Chung, S. (2019).
\newblock Melu: Meta-learned user preference estimator for cold-start
  recommendation.
\newblock In \emph{Proceedings of the SIGKDD}. 1073--1082
\bibAnnoteFile{lee2019melu}

\bibitem[{Li et~al.(2019)Li, Jing, Lu, Zhu, Yang, and Huang}]{li2019zero}
Li, J., Jing, M., Lu, K., Zhu, L., Yang, Y., and Huang, Z. (2019).
\newblock From zero-shot learning to cold-start recommendation.
\newblock In \emph{Proceedings of the AAAI}. vol.~33, 4189--4196
\bibAnnoteFile{li2019zero}

\bibitem[{Liang et~al.(2015)Liang, Zhan, and Ellis}]{liang2015content}
Liang, D., Zhan, M., and Ellis, D.~P. (2015).
\newblock Content-aware collaborative music recommendation using pre-trained
  neural networks.
\newblock In \emph{Proceedings of the ISMIR}. 295--301
\bibAnnoteFile{liang2015content}

\bibitem[{Lin et~al.(2015)Lin, Liu, Sun, Liu, and Zhu}]{lin2015transr}
Lin, Y., Liu, Z., Sun, M., Liu, Y., and Zhu, X. (2015).
\newblock Learning entity and relation embeddings for knowledge graph
  completion.
\newblock In \emph{Proceedings of the AAAI} (AAAI Press), 2181–2187
\bibAnnoteFile{lin2015transr}

\bibitem[{Liu et~al.(2020)Liu, Zhou, Zhao, Wang, Ju, Deng et~al.}]{liu2020k}
Liu, W., Zhou, P., Zhao, Z., Wang, Z., Ju, Q., Deng, H., et~al. (2020).
\newblock K-bert: Enabling language representation with knowledge graph.
\newblock In \emph{Proceedings of the AAAI}. 2901--2908
\bibAnnoteFile{liu2020k}

\bibitem[{Liu et~al.(2019)Liu, Ott, Goyal, Du, Joshi, Chen
  et~al.}]{liu2019roberta}
Liu, Y., Ott, M., Goyal, N., Du, J., Joshi, M., Chen, D., et~al. (2019).
\newblock Roberta: A robustly optimized bert pretraining approach.
\newblock \emph{arXiv preprint arXiv:1907.11692}
\bibAnnoteFile{liu2019roberta}

\bibitem[{Ma et~al.(2011)Ma, Zhou, Liu, Lyu, and King}]{ma2011recommender}
Ma, H., Zhou, D., Liu, C., Lyu, M.~R., and King, I. (2011).
\newblock Recommender systems with social regularization.
\newblock In \emph{Proceedings of the WSDM}. 287--296
\bibAnnoteFile{ma2011recommender}

\bibitem[{Manotumruksa et~al.(2016)Manotumruksa, Macdonald, and
  Ounis}]{manotumruksa2016regularising}
Manotumruksa, J., Macdonald, C., and Ounis, I. (2016).
\newblock Regularising factorised models for venue recommendation using friends
  and their comments.
\newblock In \emph{Proceedings of the CIKM}. 1981--1984
\bibAnnoteFile{manotumruksa2016regularising}

\bibitem[{McCarley(2019)}]{mccarley2019pruning}
McCarley, J.~S. (2019).
\newblock Pruning a bert-based question answering model.
\newblock \emph{arXiv preprint arXiv:1910.06360}
\bibAnnoteFile{mccarley2019pruning}

\bibitem[{McPherson et~al.(2001)McPherson, Smith-Lovin, and
  Cook}]{mcpherson2001birds}
McPherson, M., Smith-Lovin, L., and Cook, J.~M. (2001).
\newblock Birds of a feather: Homophily in social networks.
\newblock \emph{Annual review of sociology} 27, 415--444
\bibAnnoteFile{mcpherson2001birds}

\bibitem[{Mikolov et~al.(2013)Mikolov, Sutskever, Chen, Corrado, and
  Dean}]{mikolov2013distributed}
Mikolov, T., Sutskever, I., Chen, K., Corrado, G.~S., and Dean, J. (2013).
\newblock Distributed representations of words and phrases and their
  compositionality.
\newblock In \emph{Proceedings of the NIPS}. 3111--3119
\bibAnnoteFile{mikolov2013distributed}

\bibitem[{Mudrakarta et~al.(2018)Mudrakarta, Sandler, Zhmoginov, and
  Howard}]{mudrakarta2018k}
Mudrakarta, P.~K., Sandler, M., Zhmoginov, A., and Howard, A. (2018).
\newblock K for the price of 1: Parameter-efficient multi-task and transfer
  learning.
\newblock In \emph{Proceedings of the ICLR}
\bibAnnoteFile{mudrakarta2018k}

\bibitem[{Nguyen et~al.(2017)Nguyen, Wistuba, Grabocka, Drumond, and
  Schmidt-Thieme}]{nguyen2017personalized}
Nguyen, H.~T., Wistuba, M., Grabocka, J., Drumond, L.~R., and Schmidt-Thieme,
  L. (2017).
\newblock Personalized deep learning for tag recommendation.
\newblock In \emph{Proceedings of the PAKDD} (Springer), 186--197
\bibAnnoteFile{nguyen2017personalized}

\bibitem[{Ni et~al.(2018)Ni, Ou, Liu, Li, Ou, Zeng et~al.}]{ni2018perceive}
Ni, Y., Ou, D., Liu, S., Li, X., Ou, W., Zeng, A., et~al. (2018).
\newblock Perceive your users in depth: Learning universal user representations
  from multiple e-commerce tasks.
\newblock In \emph{Proceedings of the SIGKDD}. 596--605
\bibAnnoteFile{ni2018perceive}

\bibitem[{Qin et~al.(2020)Qin, Zhu, Zhuang, Guo, Zhang, Zhang
  et~al.}]{qin2020survey}
Qin, C., Zhu, H., Zhuang, F., Guo, Q., Zhang, Q., Zhang, L., et~al. (2020).
\newblock A survey on knowledge graph-based recommender systems.
\newblock \emph{SCIENTIA SINICA Informationis} 50, 937--956
\bibAnnoteFile{qin2020survey}

\bibitem[{Qiu et~al.(2020)Qiu, TianXiang, Yige, Yunfan, Ning, and
  Xuanjing}]{xipengpre}
Qiu, X., TianXiang, S., Yige, X., Yunfan, S., Ning, D., and Xuanjing, H.
  (2020).
\newblock Pre-trained models for natural language processing: A survey.
\newblock \emph{SCIENCE CHINA Technological Sciences}
\bibAnnoteFile{xipengpre}

\bibitem[{Rendle(2010)}]{rendle2010factorization}
Rendle, S. (2010).
\newblock Factorization machines.
\newblock In \emph{Proceedings of the ICDM} (IEEE), 995--1000
\bibAnnoteFile{rendle2010factorization}

\bibitem[{Rendle et~al.(2010)Rendle, Freudenthaler, and
  Schmidt-Thieme}]{rendle2010factorizing}
Rendle, S., Freudenthaler, C., and Schmidt-Thieme, L. (2010).
\newblock Factorizing personalized markov chains for next-basket
  recommendation.
\newblock In \emph{Proceedings of the WWW}. 811--820
\bibAnnoteFile{rendle2010factorizing}

\bibitem[{Rietzler et~al.(2020)Rietzler, Stabinger, Opitz, and
  Engl}]{rietzler2020adapt}
Rietzler, A., Stabinger, S., Opitz, P., and Engl, S. (2020).
\newblock Adapt or get left behind: Domain adaptation through bert language
  model finetuning for aspect-target sentiment classification.
\newblock In \emph{Proceedings of the LREC}. 4933--4941
\bibAnnoteFile{rietzler2020adapt}

\bibitem[{Sathish et~al.(2019)Sathish, Mehrotra, Dhinwa, and
  Das}]{sathish2019graph}
Sathish, V., Mehrotra, T., Dhinwa, S., and Das, B. (2019).
\newblock Graph embedding based hybrid social recommendation system.
\newblock \emph{arXiv preprint arXiv:1908.09454}
\bibAnnoteFile{sathish2019graph}

\bibitem[{Shi et~al.(2010)Shi, Larson, and Hanjalic}]{shi2010list}
Shi, Y., Larson, M., and Hanjalic, A. (2010).
\newblock List-wise learning to rank with matrix factorization for
  collaborative filtering.
\newblock In \emph{Proceedings of the RecSys}. 269--272
\bibAnnoteFile{shi2010list}

\bibitem[{Song et~al.(2016)Song, Elkahky, and He}]{song2016multi}
Song, Y., Elkahky, A.~M., and He, X. (2016).
\newblock Multi-rate deep learning for temporal recommendation.
\newblock In \emph{Proceedings of the SIGIR}. 909--912
\bibAnnoteFile{song2016multi}

\bibitem[{Stickland and Murray(2019)}]{stickland2019bert}
Stickland, A.~C. and Murray, I. (2019).
\newblock Bert and pals: Projected attention layers for efficient adaptation in
  multi-task learning.
\newblock In \emph{Proceedings of the ICML}. 5986--5995
\bibAnnoteFile{stickland2019bert}

\bibitem[{Sun et~al.(2019)Sun, Liu, Wu, Pei, Lin, Ou et~al.}]{sun2019bert4rec}
Sun, F., Liu, J., Wu, J., Pei, C., Lin, X., Ou, W., et~al. (2019).
\newblock Bert4rec: Sequential recommendation with bidirectional encoder
  representations from transformer.
\newblock In \emph{Proceedings of the CIKM}. 1441--1450
\bibAnnoteFile{sun2019bert4rec}

\bibitem[{Tan et~al.(2016)Tan, Wan, and Xiao}]{tan2016neural}
Tan, J., Wan, X., and Xiao, J. (2016).
\newblock A neural network approach to quote recommendation in writings.
\newblock In \emph{Proceedings of the CIKM}. 65--74
\bibAnnoteFile{tan2016neural}

\bibitem[{Tang et~al.(2013)Tang, Hu, and Liu}]{tang2013social}
Tang, J., Hu, X., and Liu, H. (2013).
\newblock Social recommendation: a review.
\newblock \emph{Social Network Analysis and Mining} 3, 1113--1133
\bibAnnoteFile{tang2013social}

\bibitem[{Tang and Wang(2018)}]{tang2018personalized}
Tang, J. and Wang, K. (2018).
\newblock Personalized top-n sequential recommendation via convolutional
  sequence embedding.
\newblock In \emph{Proceedings of the WSDM}. 565--573
\bibAnnoteFile{tang2018personalized}

\bibitem[{Tang et~al.(2019)Tang, Wang, Yang, and Song}]{tang2019akupm}
Tang, X., Wang, T., Yang, H., and Song, H. (2019).
\newblock Akupm: Attention-enhanced knowledge-aware user preference model for
  recommendation.
\newblock In \emph{Proceedings of the SIGKDD}. 1891--1899
\bibAnnoteFile{tang2019akupm}

\bibitem[{Vaswani et~al.(2017)Vaswani, Shazeer, Parmar, Uszkoreit, Jones, Gomez
  et~al.}]{vaswani2017attention}
Vaswani, A., Shazeer, N., Parmar, N., Uszkoreit, J., Jones, L., Gomez, A.~N.,
  et~al. (2017).
\newblock Attention is all you need.
\newblock In \emph{Proceedings of the NIPS}. 5998--6008
\bibAnnoteFile{vaswani2017attention}

\bibitem[{Wang et~al.(2018{\natexlab{a}})Wang, Zhang, Hou, Xie, Guo, and
  Liu}]{wang2018shine}
Wang, H., Zhang, F., Hou, M., Xie, X., Guo, M., and Liu, Q.
  (2018{\natexlab{a}}).
\newblock Shine: Signed heterogeneous information network embedding for
  sentiment link prediction.
\newblock In \emph{Proceedings of the WSDM}. 592--600
\bibAnnoteFile{wang2018shine}

\bibitem[{Wang et~al.(2018{\natexlab{b}})Wang, Zhang, Xie, and
  Guo}]{wang2018dkn}
Wang, H., Zhang, F., Xie, X., and Guo, M. (2018{\natexlab{b}}).
\newblock Dkn: Deep knowledge-aware network for news recommendation.
\newblock In \emph{Proceedings of the WWW}. 1835--1844
\bibAnnoteFile{wang2018dkn}

\bibitem[{Wang et~al.(2020)Wang, Tang, Duan, Wei, Huang, Cao
  et~al.}]{wang2020k}
Wang, R., Tang, D., Duan, N., Wei, Z., Huang, X., Cao, C., et~al. (2020).
\newblock K-adapter: Infusing knowledge into pre-trained models with adapters.
\newblock \emph{arXiv preprint arXiv:2002.01808}
\bibAnnoteFile{wang2020k}

\bibitem[{Wang et~al.(2014)Wang, Zhang, Feng, and Chen}]{wang2014knowledge}
Wang, Z., Zhang, J., Feng, J., and Chen, Z. (2014).
\newblock Knowledge graph embedding by translating on hyperplanes.
\newblock In \emph{Proceedings of the AAAI}. 1112--1119
\bibAnnoteFile{wang2014knowledge}

\bibitem[{Wen et~al.(2018)Wen, Guo, Chen, and Ma}]{wen2018network}
Wen, Y., Guo, L., Chen, Z., and Ma, J. (2018).
\newblock Network embedding based recommendation method in social networks.
\newblock In \emph{Proceedings of the WWW}. 11--12
\bibAnnoteFile{wen2018network}

\bibitem[{Wu et~al.(2017)Wu, Ahmed, Beutel, Smola, and Jing}]{wu2017recurrent}
Wu, C.-Y., Ahmed, A., Beutel, A., Smola, A.~J., and Jing, H. (2017).
\newblock Recurrent recommender networks.
\newblock In \emph{Proceedings of the WSDM}. 495--503
\bibAnnoteFile{wu2017recurrent}

\bibitem[{Xu et~al.(2016)Xu, Chen, Lukasiewicz, Miao, and Meng}]{xu2016tag}
Xu, Z., Chen, C., Lukasiewicz, T., Miao, Y., and Meng, X. (2016).
\newblock Tag-aware personalized recommendation using a deep-semantic
  similarity model with negative sampling.
\newblock In \emph{Proceedings of the CIKM}. 1921--1924
\bibAnnoteFile{xu2016tag}

\bibitem[{Yang et~al.(2019)Yang, Xu, Tong, Gao, Guo, and Wen}]{yang2019pre}
Yang, J., Xu, J., Tong, J., Gao, S., Guo, J., and Wen, J. (2019).
\newblock Pre-training of context-aware item representation for next basket
  recommendation.
\newblock \emph{arXiv preprint arXiv:1904.12604}
\bibAnnoteFile{yang2019pre}

\bibitem[{Yu et~al.(2016)Yu, Liu, Wu, Wang, and Tan}]{yu2016dynamic}
Yu, F., Liu, Q., Wu, S., Wang, L., and Tan, T. (2016).
\newblock A dynamic recurrent model for next basket recommendation.
\newblock In \emph{Proceedings of the SIGIR}. 729--732
\bibAnnoteFile{yu2016dynamic}

\bibitem[{Yu et~al.(2018)Yu, Gao, Li, Yin, and Liu}]{yu2018adaptive}
Yu, J., Gao, M., Li, J., Yin, H., and Liu, H. (2018).
\newblock Adaptive implicit friends identification over heterogeneous network
  for social recommendation.
\newblock In \emph{Proceedings of the CIKM}. 357--366
\bibAnnoteFile{yu2018adaptive}

\bibitem[{Yuan et~al.(2020)Yuan, He, Karatzoglou, and
  Zhang}]{yuan2020parameter}
Yuan, F., He, X., Karatzoglou, A., and Zhang, L. (2020).
\newblock Parameter-efficient transfer from sequential behaviors for user
  modeling and recommendation.
\newblock In \emph{Proceedings of the SIGIR}. 1469--1478
\bibAnnoteFile{yuan2020parameter}

\bibitem[{Zhang et~al.(2016)Zhang, Yuan, Lian, Xie, and
  Ma}]{zhang2016collaborative}
Zhang, F., Yuan, N.~J., Lian, D., Xie, X., and Ma, W.-Y. (2016).
\newblock Collaborative knowledge base embedding for recommender systems.
\newblock In \emph{Proceedings of the SIGKDD}. 353--362
\bibAnnoteFile{zhang2016collaborative}

\bibitem[{Zhang et~al.(2018)Zhang, Hu, Shi, Wu, and Wang}]{zhang2018matrix}
Zhang, M., Hu, B., Shi, C., Wu, B., and Wang, B. (2018).
\newblock Matrix factorization meets social network embedding for rating
  prediction.
\newblock In \emph{Proceedings of the APWeb-WAIM} (Springer), 121--129
\bibAnnoteFile{zhang2018matrix}

\bibitem[{Zhang et~al.(2017)Zhang, Ai, Chen, and Croft}]{zhang2017joint}
Zhang, Y., Ai, Q., Chen, X., and Croft, W.~B. (2017).
\newblock Joint representation learning for top-n recommendation with
  heterogeneous information sources.
\newblock In \emph{Proceedings of the CIKM}. 1449--1458
\bibAnnoteFile{zhang2017joint}

\bibitem[{Zhang et~al.(2019)Zhang, Han, Liu, Jiang, Sun, and
  Liu}]{zhang2019ernie}
Zhang, Z., Han, X., Liu, Z., Jiang, X., Sun, M., and Liu, Q. (2019).
\newblock Ernie: Enhanced language representation with informative entities.
\newblock In \emph{Proceedings of the ACL}. 1441--1451
\bibAnnoteFile{zhang2019ernie}

\bibitem[{Zhao et~al.(2017)Zhao, Zhang, Xia, Ding, Yin, and
  Tang}]{zhao2017deep}
Zhao, X., Zhang, L., Xia, L., Ding, Z., Yin, D., and Tang, J. (2017).
\newblock Deep reinforcement learning for list-wise recommendations.
\newblock \emph{arXiv preprint arXiv:1801.00209}
\bibAnnoteFile{zhao2017deep}

\bibitem[{Zheng et~al.(2017)Zheng, Noroozi, and Yu}]{zheng2017joint}
Zheng, L., Noroozi, V., and Yu, P.~S. (2017).
\newblock Joint deep modeling of users and items using reviews for
  recommendation.
\newblock In \emph{Proceedings of the WSDM}. 425--434
\bibAnnoteFile{zheng2017joint}

\end{thebibliography}



\end{document}